\documentclass[prl,aps,floats,twocolumn,showpacs,nofootinbib]{revtex4}

\usepackage{amssymb}
\usepackage{amsmath}
\usepackage{cancel}

\usepackage{graphicx}
\usepackage{graphics}
\usepackage{dcolumn}
\usepackage{color}
\usepackage{rotate}
\usepackage{fancyhdr} 
\usepackage{hyperref}
\usepackage{indentfirst}

\usepackage{umoline}
\usepackage{ulem}

\def\be{\begin{equation}}
\def\ee{\end{equation}}
\def\bea{\begin{eqnarray}}
\def\eea{\end{eqnarray}}
\def\ba{\begin{eqnarray}}
\def\ea{\end{eqnarray}}

\def\VEV#1{\left\langle #1 \right\rangle}

\begin{document}

\title{Statistical diagnostics to identify Galactic foregrounds in B-mode maps}

\author{Marc Kamionkowski$^{1}$ and Ely D.\ Kovetz$^{2}$}

\affiliation{$^1$Department of Physics and Astronomy, Johns
     Hopkins University, Baltimore, MD 21218, USA} 
\affiliation{$^2$Theory Group, Department of Physics and Texas
     Cosmology Center, The University of Texas at Austin, TX
     78712, USA}

\date{\today}

\begin{abstract}
Recent developments in the search for inflationary
gravitational waves in the cosmic microwave background (CMB)
polarization motivate the search for new diagnostics to
distinguish the Galactic foreground contribution to B modes from the
cosmic signal.  We show that B modes from these
foregrounds should exhibit a local hexadecapolar departure in
power from statistical isotropy (SI).  We present a simple
algorithm to search for a
uniform SI violation of this sort, as may arise in a
sufficiently small patch of sky.  We then show how to search for
these effects if the orientation of the SI violation varies
across the survey region, as is more likely to occur in surveys
with more sky coverage.  If detected, these departures
from Gaussianity would indicate some level of Galactic foreground
contamination in the B-mode maps.  Given uncertainties about 
foreground properties, though, caution should be exercised in attributing
a null detection to an absence of foregrounds.
\end{abstract}

\maketitle

The BICEP2 collaboration recently reported \cite{Ade:2014xna}
evidence for the signature \cite{Kamionkowski:1996zd} of
inflationary gravitational waves \cite{Abbott:1984fp} in the B-mode
component \cite{Kamionkowski:1996ks,Zaldarriaga:1996xe} of the
polarization of the cosmic microwave background (CMB).  The
extraodinary stream of papers \cite{attention} that have
followed this announcement provides some indication of the
significance of a B-mode detection.  However, the remarkable
implications of this measurement---the detection of a new relic
from inflation---demand that the results receive the deepest
possible scrutiny.  Discussions that have 
taken place since the March 2014 announcement indicate that more
work must be done to establish, with the type of confidence such
an extraordinary result warrants, that the B-mode signal cannot
be attributed fully to polarized emission from interstellar dust
(see, e.g., Refs.~\cite{Flauger:2014qra,Mortonson:2014bja}).

The gold standard to distinguish CMB from foregrounds (primarily
synchrotron and dust emission from the Milky Way) has typically
been to obtain high--signal-to-noise maps at multiple
frequencies.  Important steps in this direction should soon
be taken for the BICEP2 B-mode signal with new data from the
100-GHz Keck Array \cite{keckarray} and from polarization
measurements from Planck \cite{planck} at higher frequencies,
and soon indepenently with other experiments (e.g.,
Ref.~\cite{class,Fraisse:2011xz,Aiola:2012goa,Lazear:2014bga}).
However, these measurements may, like
any others, ultimately have limits.  For example,
extrapolation of measurements of the B-mode power from dust
obtained with Planck's 353-GHz channel to BICEP2's 150 GHz
channel may suffer from theoretical
uncertainties in the frequency dependence of the dust
polarization\footnote{Indeed, the frequency-dependent models 1 and 3 of Ref.~\cite{Draine:2008hu} (see Fig.~(8)) predict an opposite trend with frequency than observed (see Fig.~(13) in Ref.~\cite{Ade:2014P}), indicating our theoretical uncertainty.}. The use of spatial
cross-correlations between different frequency channels
may be imperfect if the depths in the interstellar medium probed
by those two frequencies differ.
Even if the dust contribution turns out to be small
enough that such subtleties do not prevent the confident establishment
of a gravitational-wave signal, every detail about the
early Universe that we can extract from detailed
characterization of the B-mode signal will be priceless.  It is
thus imperative that we remain ever vigilant in our quest to
find new ways to root out contaminants to the cosmic B-mode signal.

Here we propose two statistical tests that can
be performed on an observed B-mode map\footnote{In principle, one might just look in the data for a preferred orientation in the polarization map. Most CMB experiments, however, measure only differences in polarization and thus are not equipped to measure the average orientation.  Moreover, much of what we discuss below for B modes also applies to E modes, but the additional information in E modes is likely to be swamped by cosmic variance from the dominant density-perturbation contribution to E modes. Still, higher-frequency E-mode maps may be useful for constructing dust orientation templates for cross-correlation with B-mode maps.}.---either a
single-frequency map or one that has been cleaned with multifrequency
information---to help identify foreground contamination.
The idea is simple:  The departures in the inflationary
gravitational-wave signal from Gaussianity and statistical
isotropy (SI) are expected to be extremely small \cite{Maldacena:2011nz}.  Any
statistically significant departure from Gaussianity or
SI would thus indicate some non-cosmic
contamination.

The question, though, is what {\it type} of non-Gaussianity or
SI violation should we be seeking?  Here we argue that the
polarization due to foregrounds over a sufficiently small region
of the sky induces a hexadecapolar anisotropy in the B-mode
power, something that should be relatively simple to seek.  We
then show how to look for a spatially-varying SI violation of this
sort, something that is more likely to describe the foreground
polarization pattern on larger patches of sky.

Let us begin by understanding how this SI violation arises, in particular for the case of dust. 
Polarized emission from dust stems from the alignment of
spinning dust grains with the Galactic magnetic field
\cite{Draine:2008hu} (which also determines the synchrotron
polarization).  Galactic magnetic fields are
known to have long-range
correlations, implying an orientation angle that is fairly
coherent on large regions of the sky \cite{Ade:2014gna}, and
perhaps larger than the patch covered by BICEP2.
There may, of course, be significant changes in that orientation
angle in small sky patches if there are regions of high-density
plasma in the ISM in that patch.  The BICEP2 patch, however,
which lies in the ``Southern Hole," was chosen for the expectation that 
it was relatively clean \cite{Ade:2014gua} and thus likely free from rapid variation in
the orientation angle (as shown in  Fig.~(13) of Ref.~\cite{Ade:2014gna}, 
the typical angle dispersion of dust polarization is lowest in the highest 
polarization-fraction regions of the sky, those cleanest and most suitable for
 B-mode measurements). Furthermore, measurements of polarized absorption of
starlight (which is correlated with polarized dust emission
\cite{Ade:2014ina}) in the BICEP2 region may provide some
empirical indication that the orientation of the dust
polarization in the BICEP2 patch is roughly uniform, as noted by
Ref.~\cite{Flauger:2014qra}. However, as this data lies near the edges
of the field, it cannot provide a robust constraint on the entire patch. 

Let us therefore consider a B-mode signal from a map in which
the orientation angle of the polarization is constant.  
The Stokes parameters $Q(\vec{\theta})$ and
$U(\vec{\theta})$, measured as a function of position
$\vec{\theta}=(\theta_x,\theta_y)$ on a flat region of sky, are
components of a polarization tensor,
\begin{equation}
     P_{ab} = \frac{1}{\sqrt{2}}\left(
     \begin{array}{cc}
     Q(\vec{\theta})& U(\vec{\theta})\\
     U(\vec{\theta}) & -Q(\vec{\theta})\\
     \end{array}\right ).
\label{eqn:QUmatrix}
\end{equation}
The polarization map is then decomposed into scalar and
pseudoscalar components $E(\vec\theta)$ and $B(\vec\theta)$ by
\begin{equation}
\label{eq.nabla2}
     \nabla^2 E=\partial_a \partial_bP_{ab}\:\:\:;\;\:\:
     \nabla^2 B= \epsilon_{ab}\partial_a\partial_cP_{cb},
\end{equation}
where $\epsilon_{ab}$ is the antisymmetric tensor. 
The Fourier components of $E(\vec{\theta})$ and
$B(\vec{\theta})$ are 
\begin{eqnarray}
     \tilde{E}(\vec{l}) & = & 2^{-1/2} \left[ \cos
     2\varphi_{\vec l} \tilde{Q}(\vec{l}) + \sin2\varphi_{\vec
     l} \tilde{U}(\vec{l}) \right],\\
     \tilde{B}(\vec{l}) & = & 2^{-1/2}
     \left[- \sin 2\varphi_{\vec l} \tilde{Q}(\vec{l})+\cos
     2\varphi_{\vec l} \tilde{U}(\vec{l}) \right],
\label{eqn:GCFouriercomponents}
\end{eqnarray}
in terms of the Fourier transforms $\tilde Q(\vec l)$ and
$\tilde U(\vec l)$ of the Stokes parameters and the angle
$\varphi_{\vec l}$ that $\vec l$ makes with the $\theta_x$ axis.

If the polarization is constant across the map with orientation
$\alpha=(1/2)\arctan(U/Q)$ with respect to the $\theta_x$
axis, then the Fourier modes for E and B will be,
\begin{eqnarray}
     \tilde{E}(\vec{l}) & = & \frac{\tilde P(\vec l)}{\sqrt{2}}
     \cos\left[2(\alpha-\varphi_{\vec l})\right] ,\\
     \tilde{B}(\vec{l}) & = & \frac{\tilde P(\vec l)}{\sqrt{2}}
     \sin\left[2(\alpha-\varphi_{\vec l})\right] ,
\label{eqn:GCFouriercomponentsrorated}
\end{eqnarray}
where $\tilde P(\vec l)$
is the Fourier transform of the polarization amplitude
$P(\vec\theta)\equiv (Q^2+U^2)^{1/2}(\vec\theta)$.  We thus see
that {\it if the orientation angle of polarization is constant,
the B modes that result are not statistically isotropic}.  They
are, rather, modulated by $\sin\left[2\left(\alpha-\varphi_{\vec l}
\right)\right]$.

An estimator for this departure from statistical isotropy in the
B-mode map can be obtained through a straightforward
augmentation of the usual algorithm to determine the amplitude of the 
B-mode power.  Eq.~(\ref{eqn:GCFouriercomponentsrorated})---which
is what we expect if the observed B modes are due entirely to
dust and if the dust polarization has uniform
orientation---implies that the mean-square amplitude of each
B-mode coefficient is,
\begin{equation}
     \VEV{ \left| \tilde B(\vec l) \right|^2} = A C_l^f \left[ 1 -
     \cos 4 \alpha \cos 4 \varphi_{\vec l} - \sin 4\alpha \sin
     4\varphi_{\vec l} \right],
\label{eqn:modulation}
\end{equation}
where $C_l^f$ parametrizes an assumed fiducial $l$ dependence
(e.g., $C_l^f\propto l^{-2.22}$, as current measurements suggest
\cite{Ade:2014gna})
and $A$ an amplitude of the signal.  Note that although the
modulation of the Fourier amplitudes is quadrupolar ($\propto
e^{2i\alpha}$), the departure from statistical isotropy in the
power spectrum is a hexadecapole; it has an $e^{4i\alpha}$
dependence.

More generally, if the orientation of the dust polarization is
not perfectly uniform, but is rather spread over some small
range $\delta\alpha$, then the modulation in
Eq.~(\ref{eqn:modulation}) will be reduced by a factor $\sim
(\delta\alpha)/\alpha$.  Thus, to test for dust, we should aim
to measure the parameters in the angle-dependent power spectrum,
\begin{equation}
     \VEV{ \left| \tilde B(\vec l) \right|^2} = A C_l^f \left[ 1 -
     f_c \cos 4 \varphi_{\vec l} - f_s \sin 4\varphi_{\vec l}
     \right],
\label{angledepPS}
\end{equation}
where $f_s,f_c<1$ measure the departure from statistical
isotropy, and the dust-polarization orientation, if these
parameters are found to be nonzero, is $\alpha =
(1/4)\arctan(f_s/f_c)$.

The minimum-variance estimator for the isotropic amplitude $A$
is the usual one,
\begin{equation}
     \widehat{A} = \frac{ \sum_{\vec l} \left|\tilde
     B_{\vec l} \right|^2 C_l^f/\sigma_l^2} { \sum_{\vec l}
     \left(C_l^f \right)^2/\sigma_l^2},
\end{equation}
where the sum is over all Fourier modes $\vec l$ with amplitudes
$\tilde B(\vec l)$ each measured with variance $\sigma_l^2$
(which may receive contributions from detector noise and from
lensing-induced B modes \cite{Zaldarriaga:1998ar,Ade:2014afa}).
The minimum-variance estimators for the amplitudes of the SI-violating 
terms are likewise,
\begin{equation}
     \widehat{A{ f_c}} = \frac{ \sum_{\vec l} \left|\tilde
     B_{\vec l} \right|^2 C_l^f \cos 4\varphi_{\vec
     l}/\sigma_l^2} { \sum_{\vec l} \left(C_l^f \cos
     4\varphi_{\vec l}\right)^2/\sigma_l^2},
\end{equation}
and similarly for the $f_s$ term with the replacement $\cos\to\sin$.  If
there is no prior information about the orientation of the dust
polarization, then the parameters $f_s$ and $f_c$ are both
obtained simultaneously and independently from the data.  If,
however, there is
some prior information about the expected orientation---e.g.,
from starlight polarization---then the ratio $f_s/f_c$ can be
fixed and the sensitivity to dust-induced SI violation thus
accordingly improved.  Either way, {\it any statistically
significant detection of nonzero $f_s$ and/or $f_c$ indicates at
least some contamination of the cosmic signal.}  
If, moreover, either of the inferred values $f_c$ or
$f_s$ differs significantly from zero, then there is good
evidence that the signal is predominantly non-cosmic.  If there
is strong reason to believe that the foreground-polarization
orientation is indeed uniform across the survey,
then a strong null result may imply that the observed signal is
{\it not} foreground dominated.  If, though, that orientation is
uncertain, then a null result in this SI test cannot
be used to rule out foreground contamination.

The variances and covariances with which the
parameters $A$, $f_s$, and $f_c$ can be measured are easily
derived.  However, they will depend considerably on the details of
any given experiment and perhaps a bit on the
fact that the lensing-induced B-mode map is not precisely
Gaussian.  We thus leave these covariances
to simulations of the complete analysis pipelines.
Heuristically, though, the estimator measures the difference in
the B-mode power for modes oriented perpendicular/parallel to
some axis versus those oriented at $45^\circ$.  If there is a
$\gtrsim5\sigma$ detection of power, and if that power is due
entirely to uniformly oriented dust, then the violation of
statistical isotropy should appear with high statistical
significance. 
Indeed, a crude estimate for the minimum amplitude $A$ that 
can be measured at 1$\sigma$ is given by
\begin{eqnarray}
&&\sigma_{\widehat{A}}^{-2}={\sum_{\vec l}
     \left(C_l^f \right)^2/\sigma_l^2}\sim\Omega{\int \frac{d^2l}{(2\pi)^2}
     \left(C_l^f \right)^2/\sigma_l^2} \nonumber \\
&&  = 4\pi f_{\rm sky}{\int \frac{d^2l}{(2\pi)^2}
     \left(C_l^f \right)^2/\sigma_l^2} = 
    2f_{\rm sky}{\int ldl
    \left(C_l^f \right)^2/\sigma_l^2}
    \nonumber \\
\end{eqnarray}
and $\sigma_{\widehat{Af_c}}^{2}=2\sigma_{\widehat{A}}^{2}$ (as $\int_0^{2\pi}d\varphi=2\int_0^{2\pi}\cos^2(4\varphi)d\varphi=2\pi$).
The signal-to-noise in a particular experiment is governed by the sensitivity per Fourier mode
\be
  \sigma_l = \sqrt{\frac{2}{ f_{\rm
  sky}(2l+1)}}\left(C^{\rm lens}_l +
  f_{\rm sky} w^{-1}(T) e^{l^2\sigma_b^2}\right),
\ee 
where the pixel noise $\sigma_{\rm pix}=s/\sqrt{t_{\rm pix}}$ is determined by the detector sensitivity $s$ and the observation time $t_{\rm pix}=T/N_{\rm pix}$ dedicated to each pixel, and where we used the definition $w^{-1}(T)\equiv4\pi s^2/T$. 

It should be noted that some of the dust-polarization templates
used by BICEP2 and investigated in subsequent work were
constructed assuming a uniform dust-polarization orientation.
The departures from SI considered above are then effectively
incorporated into the data-template cross-correlation analyses
done already. Those cross-correlations, though, may still
vanish if either (1) the assumed orientation angle is incorrect, 
or (2) the spatial variation of the polarization
amplitude is not correctly represented, as can be seen in Ref.~\cite{Flauger:2014qra}).
The SI-violation analysis suggested above, though, does not rely on prior
knowledge of the spatial variation of the amplitude nor the
assumed orientation angle.

So far we have supposed that the sky patch is small enough that
a uniform dust-polarization orientation may be reasonably
hypothesized.  However, future experiments will cover larger
regions of the sky (e.g.,
Ref.~\cite{class,Fraisse:2011xz,Lazear:2014bga}), and it is
increasingly likely that the foreground-polarization orientation
will meander across the survey region as the size of that region
increases.  The foreground polarization may thus be modeled in
terms of an amplitude that has rapid small-scale variation with
an orientation that has longer-range correlations.  This can be
sought in a straightforward fashion by simply measuring the
correlations in the polarization amplitude and in the
orientation angle.  If the signal is cosmic, the correlations in
both should be similar.  Evidence that those two correlation
lengths differ could indicate a non-cosmic source of
contamination.  Such an analysis, though, will likely be limited
by cosmic variance from the dominant
density-perturbation--induced polarization.

Instead, we now spell out a diagnostic for spatial variations of
the type of SI-violation above that parallels algorithms
developed to search for spatially-varying cosmic birefringence \cite{cb},
optical depth (``patchy screening'') \cite{patchy}, and
cosmological parameters \cite{Grin:2011nk}, and before those, weak
lensing \cite{lensing} (which has now been detected
\cite{Ade:2014afa,lensingdetection}).  For clarity, we work here in the
flat-sky limit; the generalization to the full sky is
straightforward and follows this other previous analogous work.

We suppose that there are variations of the orientation angle
that vary slowly across the sky with small-scale fluctuations in
the polarization amplitude.  We thus assume the polarization can
be written,
\begin{equation}
     P_{ab}(\vec\theta)=P_{ab}^o(\vec\theta)\phi(\vec\theta)
\label{eqn:decomposition}
\end{equation}
in terms of a smooth  ``orientation field'' $P_{ab}^o(\theta)$ with
Stokes parameters $Q_o(\vec\theta)$ and $U_o(\vec\theta)$ and
a more rapidly-varying polarization-amplitude field
$\phi(\vec\theta)$ (which for dust should be correlated with the
dust-intensity field, although we do not use any such information
here).  The orientation field can be
decomposed in the usual manner into E and B modes
$E_o(\vec\theta)$ and $B_o(\vec\theta)$.  There is an ambiguity
in the definitions of $P_{ab}^o(\vec\theta)$ and
$\phi(\vec\theta)$---one can be increased while the other is
reduced without changing $P_{ab}$---that can be removed by
demanding, e.g., that the polarization amplitude field have unit variance
or some specific maximum value.

Consider a spatial variation of the orientation that
consists of a single Fourier mode of wavevector $\vec L$
of either the E type or the B type.  The orientation pattern in
the first case always has only nonzero $Q$ (measured with
respect to axes aligned with $\vec L$) and in the latter case
only nonzero $U$.  Thus, in the first case (E-mode orientation),
the polarization is always aligned/perpendicular to $\vec L$,
and in the second (B-mode orientation), the polarization is
always aligned at axes rotated by $45^\circ$ from $\vec L$.
Therefore, in either case---a pure-E orientation or a pure-B
orientation---the orientation of the SI violation in the
polarization $B$ modes are everywhere the same, even though the
orientation angle is changing.  Thus, in either of these two
cases, there will be SI violation in the observed B modes that
is uniform across the sky, and the simple SI-violation test
above will capture the effect in its entirety and have a positive result.

To make things a bit more interesting, consider an orientation
that rotates clockwise as we move in the $\theta_x$ direction,
completing a full revolution after a distance $\theta_x =
2\pi/L$.  I.e., 
\begin{equation}
     \left( \begin{array}{c} Q_o  \\ U_o \end{array}\right)
     (\vec\theta) = R_{\vec L}  \left( \begin{array}{c} \cos L \theta_x
     \\ \sin L \theta_x \end{array}\right).
\label{eqn:one}
\end{equation}
This is a linear combination of an E mode and a B mode, both of
the same $\vec L$, added out of phase---i.e., $E+iB$---and
$R_{\vec L}$ is the amplitude of this Fourier mode.  More
precisely,
\begin{equation}
     \left( \begin{array}{c} Q_o  \\ U_o \end{array}\right)
     (\vec\theta)  = \left[\frac{R_{\vec L}}{\sqrt{2}}
     \left( \begin{array}{c} 1 \\ i \end{array}\right) e^{i \vec
     L\cdot \vec \theta} + {\rm cc} \right],
\end{equation}
where now we have allowed $R_{\vec L}$ to be complex to allow a
phase different from that in Eq.~(\ref{eqn:one}).  We then
suppose that the
observed polarization is obtained by multiplying this
slowly-varying orientation field with a rapidly-varying
amplitude $\phi(\vec\theta)$; i.e.,
\begin{equation}
     \left( \begin{array}{c} Q  \\ U \end{array}\right)
     (\vec\theta) = \left[\frac{R_{\vec L}}{\sqrt{2}}
     \left( \begin{array}{c} 1 \\ i \end{array}\right) e^{i \vec
     L\cdot \vec \theta} + {\rm cc} \right] \phi(\vec\theta).
\label{eqn:rotation}
\end{equation}
Since the orientation varies over all possible values, the
observed B modes will be statistically isotropic when averaged
over the whole field, and the SI-violation test suggested above
will give a null result.  Still, the observed B modes will
exhibit {\it local} departures from SI.

We now explain how to detect this position-dependent local SI violation.
The polarization pattern in Eq.~(\ref{eqn:rotation})
yields B modes,
\begin{equation}
     \tilde B(\vec l) = \frac{i}{2} \left[ R_{\vec L}
     \tilde\phi(\vec l -\vec L) e^{2i\varphi_{\vec l}} - R_{\vec L}^*
     \tilde\phi(\vec l + \vec L) e^{-2 i \varphi_{\vec l}}\right].
\end{equation}
Before proceeding, recall that the B modes due to inflationary
gravitational waves are expected to be Gaussian and
statistically isotropic which implies that $\VEV{\tilde B(\vec
l) \tilde B^*(\vec l')} =0$ for $\vec l \neq \vec l'$.  However, we
now find that the polarization pattern in
Eq.~(\ref{eqn:rotation}) has expectation values,
\begin{eqnarray}
    \VEV{\tilde B(\vec l) \tilde B^*(\vec l')} &=& \frac{1}{4}
    \left[ |R_{\vec L}|^2 (C_{|\vec l-\vec L|}^\phi +C_{|\vec
    l+\vec L|}^\phi) \delta_{\vec l,\vec l'}  \right. \nonumber \\
    & & \left. - (R_{\vec {L}}^*)^2
    C_{|\vec l+\vec L|}^\phi e^{-2i(\varphi_{\vec
    l}+\varphi_{\vec l'})} \delta_{\vec l',\vec l + 2 \vec L} \right.
    \nonumber \\
    & & \left. - (R_{\vec {L}})^2 C_{|\vec l-\vec L|}^\phi e^{2i(\varphi_{\vec
    l}+\varphi_{\vec l'})} \delta_{\vec l',\vec l - 2 \vec
    L} \right],\nonumber\\
\label{eqn:result}
\end{eqnarray}
where  $C_l^\phi$ is the power spectrum of the modulation field
$\phi(\vec\theta)$, and $\delta_{\vec l,\vec l'}$ is shorthand for $(2\pi)^3
\delta_D(\vec l-\vec l')$, the Dirac delta function.  The first
term in Eq.~(\ref{eqn:result}), the only one that is
nonvanishing for $\vec l = \vec l'$, provides the
(angle-averaged) B-mode power spectrum for the map.  Roughly
speaking, it is the amplitude power spectrum $C_l^\phi$ smeared
in $l$ space by $L$.  As argued above, this first term indicates
that there is no departure from statistical isotropy when power is
averaged over the entire map.

The second two terms in Eq.~(\ref{eqn:result}), though, describe
the local SI violation of a polarization field due to the
small-scale modulation of a longer-range orientation field.
They indicate a
cross-correlation of a Fourier mode of wavevector $\vec l$ with
those of wavevectors $\vec l'=\vec l\pm 2\vec L$.  The
appearance of $2\vec L$ (rather than just $\vec L$) is related to the 
hexadecapolar nature of the power asymmetry.

Eq.~(\ref{eqn:result}) implies that each pair of Fourier
amplitudes $\tilde B(\vec l)$ and $\tilde B(\vec l')$ with $\vec l
- \vec l' =2 \vec L$ provides an estimator,
\begin{equation}
     \widehat{(R_{\vec L}^*)^2} =-4 \frac{\tilde B(\vec l) \tilde
     B^*(\vec l') e^{2i (\varphi_{\vec l}+\varphi_{\vec
     l'})}}{C_{|\vec l +\vec L|}^\phi},
\end{equation}
for the Fourier amplitude $R_{\vec L}^*$ (or actually, its
square) of the orientation amplitude.  One then adds the
estimators from each such $\vec l,\vec l'$ pair with
inverse-variance weighting to obtain the optimal estimator for
$(R_{\vec L}^*)^2$.  The procedure is directly analogous to that
for weak-lensing, cosmic-birefringence, and patchy-screening
reconstruction, and we leave the details to be presented
elsewhere.

{\it If any $R_{\vec L}$ (for any wavevector $\vec L$ that can be
accessed with the map) is found to be nonzero with statistical
significance, it indicates a likely contamination from foreground.} 
Naturally, when searching for deviation from SI in multiple independent $L$ modes, 
the ``look elsewhere effect'' must be properly taken into account. 
It should be possible, however, in a map that covers a
sufficiently large region of sky with sufficient signal to
noise, to measure a large number of amplitudes for $E+iB$ and
$E-iB$ modes and thus to reconstruct the orientation-angle map
$P_{ab}^o(\vec\theta)$ as a function of position on the sky.

Reassuringly, in the limit $L\to0$, where the orientation angle
becomes uniform (and taking $R_{\vec L}$ to be real, so that the orientation is aligned with $\theta_x$), Eq.~(\ref{eqn:result}) simplifies to,
\begin{equation}
    \VEV{\tilde B(\vec l) \tilde B^*(\vec l')} =
    \frac{R_{\vec {L}}^2}{2} {C^{\phi}_l} (1 - \cos 4 \varphi_{\vec l}) \delta_{\vec
    l,\vec l'},
\end{equation}
recovering the expected hexadecapolar power anisotropy.

To conclude, we have argued that polarization from dust is
likely to give rise to non-Gaussianity in the B modes they
induce, that appears as a local hexadecapolar departure from
statistical isotropy.  A simple test that will seek this SI
violation in the event that the orientation of the dust-induced
polarization is roughly constant was presented.  We also
showed how an orientation that varies across the survey region
can be sought.  Here we have only sketched out how these tests
can be done.  Much more work will be needed before they
are implemented in real data.  This will include the full
development of the optimal estimators, full-sky formalisms,
tools to deal with imperfect sky coverage, etc.  
Still, these developments should parallel the analogous
developments for, e.g., weak lensing.  The estimators for the
effects we deal with here differ in detail from those, e.g., for
weak lensing (here we seek a local hexadecapolar SI violation,
while lensing induces a quadrupolar effect), but some thought
should be given to possible confusion in a low--signal-to-noise
scenario.

We do not advocate that the foreground diagnostics we discuss
here replace multifrequency component separation.  Rather, they
can be implemented in the event of limited multifrequency
information or, in the event that multifrequency maps uncover a
cosmic signal, as a way to check for consistency or identify
residual foreground contamination in the maps.

\smallskip

We thank Sam Gralla and Hirosi Ooguri for useful
discussions and the Aspen Center for Physics for hospitality. We also 
thank the anonymous referees for very useful comments which helped
improve the clarity of this letter.
MK was supported by the John Templeton Foundation, the Simons
Foundation, and NSF grant PHY-1214000, and EDK by NSF grant
PHY-1316033.

\end{document}